\documentclass[12pt,a4]{article}
\usepackage[applemac]{inputenc}  
\usepackage{amssymb,amsmath,amsthm,amsfonts,amscd,amstext}
\usepackage[mathscr]{eucal}
\usepackage{array}
\usepackage{graphicx}

\begin{document}
\begin{titlepage}
\vspace*{7mm}

\begin{center}
{\bf \Large The Mesoscopic category, Automata and \\
Tropical Geometry} \\
\vspace*{8mm}

{Roland~Friedrich$^\mathrm{a}$  and   
Tsuyoshi~Kato$^\mathrm{b}$ 
} \\

\vspace*{3mm}

${}^\mathrm{a}$
{\em Humboldt-University Berlin} \\
\vspace{2mm}
${}^\mathrm{b}$
{\em Department of Mathematics\\ Kyoto University} \\
\vspace*{6mm}

\end{center}

\begin{abstract}
We start with comparisons of hierarchies in Biology and relate it to Quantum Field Theories. Thereby we discover many similarities and translate them into rich mathematical correspondences. The basic connection goes via scale transformations and Tropical geometry.

One of our core observations is that Cellular Automata can be naturally introduced in many physical models and refer to a (generalised) mesoscopic scale, as we point it out for the fundamental relation with Gauge Field Theories.

To illustrate our framework further, we apply it to the Witten-Kontsevich model and the Miller-Morita-Mumford classes, which in our picture arise from a dynamical system. \end{abstract}
\vfill

\begin{tabular}{ll}
{\em MSC 2000:}   &
81T40 
\\
{\em Keywords:}  & Molecular Biology; Automata, Quantum Field Theory \\
 & \\
{\em Email:}    &  {\tt rolandf@mathematik.hu-berlin.de, tkato@math.kyoto-u.ac.jp}
\end{tabular}
\end{titlepage}

\section{Hierarchies in physics and biology and structural similarities}
One of the most intriguing questions is, how can simple constituents generate complex structures or patterns and how are they mathematically described. 

In the {\bf physical domain} a prominent example is given by particles, e.g. atoms, which serve the role of the basic building blocks, to yield compound  objects, e.g. solids or liquids.

Usually one does not only make the distinction between basic constituents and complex compound systems, but one also associate characteristic scales (e.g. energy, length, time) to the objects themselves where they arise. In the example of atoms, one speaks about the micro  as compared to the macro world. Guided by experience, the mathematics which is capable to describe the objects at the different scales, also changed over time. Probably the most prominent example in this direction is the transition from classical to quantum mechanics. 

However, one also considers an intermediate range, usually termed {\bf mesoscopic}, 
which is situated ``in-between", both from the metric point of view but also from the mathematical description.

But, as mesoscopic structures build a bridge between microscopic and macroscopic objects, mesoscopic systems inherit characteristics of both worlds, e.g. classical and quantum signatures.

Mathematically, in the situation of quantum mechanics and classical mechanics, a ``semi-classical" description or a ``WKB approximation" can be applied. 
Other possible ways to describe this transitional regime are the path integral formalism, with a scaling parameter, e.g. $\hbar$, but also random matrices.

{\bf Molecular biology} is considered to be a fundamental theory to analyse many of the biological phenomena occurring at, e.g. the cellular level. The method itself is based on interaction systems among a set of words,  build on an alphabet of four letters, the DNA, and the mathematics involved is mainly of an algorithmic and combinatorial nature.

A particular aim is to understand how from such a mechanism macroscopic patterns arise, as e.g. the stripes of the skin of fishes.

In principle, these should be explicable from a physical point of view by molecular interaction systems and their electrons, but nevertheless 
it is mysterious that even though the microscopic behaviour appears to be  random, after several interaction steps at various time scales, macroscopic patterns emerge.
Henceforth,  we shall refer to such  systems of dynamical structures as the {\em dynamical topology} of {\em scale transformations}. 

From a dynamical point of view, there are three distinguishable hierarchical steps, micro, mesoscopic and macro.
However, compared with the usual physical situation, as described at the beginning, there are differences concerning the mathematical aspects of the dynamics. Whereas in physics differential equations are used as basic mathematical tools, for biology the automaton is fundamental. 

However, what is common to both is that one finds a {\em relative structure}, i.e., the micro-mesoscopic-macro hierarchies.

As we have already pointed out,  a  {\bf structural similarity} exists between biological and physical systems.

One of the common  features of their respective systems is
that both  consist of hierarchies of systems of different sizes.
The sizes of these hierarchies of three types in
biology and physics are mutually very different.
In many interesting cases the sizes of the biological hierarchies are much bigger than those of the physical one.
So the intrinsic nature of the individual materials will be different from each other.
What is similar, is the global structure of these systems, i.e., the
mechanisms of creating such hierarchies.
Such basic hierarchies in both biology and physics,
contain many refined ones.
In order to compare such biological and physical hierarchies
systematically,
it will be natural to categorise these structures, with 
the similarities interpreted as the existence of specific functors
between them.
So, a particularly important goal for us is to construct 
a category of biological hierarchies, but also the associated functor 
between the physical and biological realm:
$$\text{ Category of biological hierarchies}  \Leftrightarrow
\text{Category of physical hierarchies}.$$
One of the conclusions from our claim of the existence
of a functor between biological and physical hierarchies 
is that many physical systems should posses reasonable
discrete, or characteristics of a computer (cf.~\cite{Fr}).
This is the beginning of our program to discretise  
several physical systems, guided by the comparison with the biological side.

So, in later sections, we shall study some concrete comparisons, based on mathematical realisations, coming from both physics and biology.

\section{Mesoscopic models}
We shall describe now two models which are mesoscopic by their nature, however one of them is usually not primarily seen from this perspective. But looked at it this way one will immediately be rewarded mathematically. 
   
{\bf (A) Gauge Field Theories:}
A category of physical objects that naturally fits into our general framework are Gauge Field Theories and instantons.

As it is known~\cite{KW, P}, gauge theories can be discretised yielding lattice gauge field theories, which, as e.g. at the sub-nuclear level in Mesoscopic QCD~(\cite{LW}) posses also a mesoscopic regime.

Further, the instantons corresponds to field configurations that locally minimise the classical action, with the action functional decomposing the path space it into topological sectors, corresponding to topological charges.

Now, in the semi-classical description, the wave function can be expressed by use of the Maslov indices of its trajectories under the
constrains on the Lagrangian submanifolds (\cite{A}).
It is understood as  a {\bf Morse theory on the path spaces} $\Omega(M,p,q)$
which are infinite dimensional 
over Riemannian manifolds.
By taking the energy functional as a Morse function,
one can trace a parallel theory to the finite dimensional case.
The Morse indices tell us the structure of the homotopy types of the path spaces
 $\Omega(M,p,q)$, and it gives information on
  the topology of the underlying manifold $M$ (see \cite{M}). 

Let $M$ be a closed manifold and $f: M \to {\mathbb R}$ be a Morse function on it.
To each critical point, the degree of the Hessian is assigned,
and it is called the Morse index. In fact it is a special case of the Maslov index.
Now, Witten's complex $(C(f)_*, \partial)$ is a chain complex whose chains of degree $k$ are
generated by the sets of critical points of Morse indices $k$.
The connecting orbits are given by  the gradient flow of $f$,
which yields the boundary operators for the complex.
Its  homology is canonically isomorphic to the ordinary homology $H_*(M:{\mathbb Z})$.
This is a striking point where Morse functions connect dynamical structures with topology,
passing through the Maslov indices.

One has a loop space version of this, which is called  {\bf symplectic Floer homology}~\cite{F}.
Let $(M, \omega)$ be a symplectic manifold, and
choose a compatible almost complex structure $J$ on $(M,\omega)$.
On the loop space $C^{\infty}(S^1, M)$,
one obtains an action functional $\phi_H: C^{\infty}(S^1, M) \to {\mathbb R}$
when $\pi_2(M)=0$, by use of  a Hamiltonian function.
Now, the action functional $\phi_H$ plays the same role as the Morse function  above,
and so one obtains Floer's chain complex $(C(M,H)_*, \partial)$
on the loop space, which admits  richer structures than just homology groups.
The critical points of the action are the periodic orbits, and 
for each periodic orbit, one can associate the degree  by the Maslov index.
Thus  as the case of Morse functions on manifolds,
one can construct the Witten complex, and obtain the Floer
homology $HF(M, H, J)$.
When one forgets the extra structure mentioned above and
regards its homology $HF_*(M,H,J)$ as abelian groups, then 
it is canonically isomorphic to the singular homology $H_*(M; {\mathbb Z}_2)$ of $M$.
Thus one obtains semi-classical and classical objects from the forgetful functor:
$HF_*(M,H,J) \to H_*(M; {\mathbb Z}_2)$.

{\bf (B) Iteration dynamics for families:}
In order to understand the mechanisms of biological systems,
we propose to
express such phenomena as  iterations  of families of  maps,
and to interpret them as microscopic orbitals.
So, let us take two interval maps $f_0,f_1:[0,1] \to [0,1]$, and denote their  iterations as $O_{f_i} (x)  \equiv  \{f_i^n(x)\}_{n=0}^{\infty}$, for $i=0,1$.
We may regard them as oscillations of a molecule, and therefore we can formulate the interaction
between $O_{f_0}$ and $O_{f_1}$ in the following way: 
let $X_2=\{(k_0,k_1, \dots): k_i \in \{0,1\}\}$ be the one-sided full shift with
alphabet $\{0,1\}$, and
 choose any element $\bar{k} =(k_0,k_1, \dots) \in X_2$. 
Then, with respect to $\bar{k} \in X_2$,
 we define the interaction between $O_{f_0}$ and $O_{f_1}$ 
as a family of interval maps
$\{h^n : [0,1] \to [0,1]\}_{n=0}^{\infty}$ where:
$$h^l(x) = f_{k_l} \circ f_{k_{l-1}} \circ \dots \circ f_{k_0}(x), \quad l=0,1, \dots~.$$

Although the family of maps $\{h^n\}$ happens to behave in a complicated manner, by discarding some of the information via a projection map, patterns can be observed to emerge. 
So, let $\pi: [0,1]  \backslash \frac{1}{2} \to \{0,1\}$ 
be the projection defined as $\pi([0,\frac{1}{2}) )= 0$
and $\pi((\frac{1}{2}, 1])= 1$ and then consider the assignments for a.e. $x \in [0,1]$:
$$\bar{k}' \equiv
\pi(h^0(x), h^1(x), \dots)) \equiv (\pi \circ h^0(x), \pi \circ h^1(x), \dots) \in X_2.$$ 
Thus for each $\bar{k}$, one associates another element $\bar{k}' \in X_2$,
which we call the {\bf  interaction map}:
$$\Phi(x) \equiv \Phi(x,f_0,f_1) : X_2 \to X_2.$$
Their time scales are much larger compared to the 
$n$ of the oscillation $\{h^n\}$, as they can be determined only after knowing all the values $\pi \circ h^n(x)$ up to $n = \infty$. So,  according to our definition, this is a mesoscopic scale.

Now, if we write $\Phi(x)(k_0,k_1, \dots)=(k_0',k_1', \dots)$, for $k_i,k_j'  \in \{0, 1\}$, then although  $k_j'$ is in general determined from the data $k_0,k_1, \dots, k_j$,
 in some cases it is already given by knowing only the data $k_{j-m}, k_{j-m+1}, \dots, k_j$
 for some $m$ independent of $j$.
 This property turns out to be quite unstable and will be broken even
  for small perturbations of the original maps $f_0,f_1$.
As  such a dynamics corresponds to an automaton, 
 the original interaction can be regarded as a deformation of automata,
 whose mathematical  structure is essentially discrete and  finite.

As next we will show the relation of $\Phi$ with integrable systems. In fact,  to a variant of the above interaction maps and  
interval maps, the associated
flow $\{ \Phi(x)^t(\bar{k}) \}_{t=0}^{\infty}$ yields a  solution
of the Lotka-Volterra cell automaton, which arises from 
a new method to obtain a scaling limit, namely {\em Tropical geometry}. This geometrical theory permits to reduce discrete dynamics to cellular automata, passing through specific scaling limits.

Specifically, Tropical geometry associates  rational functions parametrised  by $t$ to $(\max, +)$-functions,
which are piece-wise linear functions on ${\mathbb R}^N$ and 
it transforms the dynamics  of the original PL function into a  complex, 
 parametrised dynamics on the affine algebraic varieties $V_t \subset {\mathbb C}^N$.

As the scaling parameter tends to infinity, the phase spaces of the dynamical systems have to be extended to ${\mathbb C}P^N \supset {\mathbb C}^N$.
At $t= \infty$, the dynamics on 
 ${\mathbb C}P^{N-1} ={\mathbb C}P^N \backslash {\mathbb C}^N$ looses the detailed  information it had for finite values of $t$.
  
Given the fact that Tropical geometry connects integrable systems, described by PDEs, with cellular automata, our perturbation via interactions of maps will allow us to extend such a  connection to `neighbourhoods'  of  the dynamics, including non-integrable systems~\cite{K6}. 

Now, for the hierarchies of scaling limits, i.e.,
$$\begin{matrix}
& \text{Micro} &  \Rightarrow & \text{Mesoscopic}  & \Rightarrow & \text{Macro} \\
& \{f_0^n(x)\}_n \leftrightarrow \{f_1^m(y)\}_m  & \Rightarrow & \text{Automaton}
 & \Rightarrow & \text{Solitons}~,
\end{matrix}$$

the interaction given by  a
piecewise linear map on ${\mathbb R}^3$, reduces to a cellular automaton 
 $\{A: V_1+ \max(0, V_2+V_3) = V_2 +   \max(0, V_1+V_4) \}$, which can 
be transformed, again by Tropical geometry, into a rational dynamics on the three dimensional affine algebraic variety
$V=\{(z_1,z_2,z_3,z_4): z_2 +z_1z_2z_4 = z_1+z_1z_2z_3\} \subset {\mathbb C}^4$.
Then, as it is shown in~\cite{K5}, one can transform the above dynamics into a KdV flow, by applying Hirota's classical method~\cite{hirota} for going from the discrete KdV to the continuous KdV equation.

Now, let us discuss  
{\bf mesocopic aspects} of Tropical geometry in a wider context.

A particular phenomenon in biology is that 
among scaling parameters from finite values to infinity, 
many biological systems work `quite' stably on
 an intermediate range between small values and $\infty$, 
 and even under changes of it. But at $t= \infty$, such systems 
 will change very differently.
For temperature many biological systems will break down at $t= \infty$.
 For hierarchies of biological systems,
scaling limits create macro patterns given by
very different dynamics from molecular interactions.

Now, let $\varphi $ be a piecewise linear map given by a $(\max,+)$-function.
Then Tropical geometry accociates with it  parametrised rational maps
$f_t$,   $t \in (1, \infty)$, which are related to $\varphi$ by 
$f_t = \log_t^{-1} \circ \varphi_t \circ \text{Log}_t$,
where the $\varphi_t$ are approximations of $\varphi$, for 
$\lim_{t \to \infty} \varphi_t =\varphi$. The correspondence between $\varphi$ and $f_t$ is one-to-one for their respective presentations.

We shall now give the following examples as illustrations: 
(1) Let us consider 
the $(\max, +)$-function 
$\varphi(x,y):= \max( \max(0, y) -x, -x) $ which corresponds to a 
 rational function
$g(z,w):= \frac{2+ w}{z}$ and which happens to be time-independent.

The dymanics given by $x_n =\varphi(x_{n-2}, x_{n-1})$ 
is  for any initial value recursive with period  $5$.
Thus one might expect that another dymanics, 
$z_n = g(z_{n-2},z_{n-1})$, will be the same,
since two dynamics given by $g$ and $\varphi_t$ are mutually conjugate 
and the limit $\lim_{t \to \infty} \varphi_t =\varphi$ holds.
However as a straight-forward calculation shows, this is not the case~\cite{K5}.

Mathematically, such phenomena occur, because
the limit $t = \infty$ does not exist for the conjugate maps Log$_t$, 
and so the passages between these dynamics  live only for a finite time $t$.
On the other hand the
dynamics given by $\varphi_t$ and $g$ are conjugate if $t < \infty$, i.e. for finite values, 
which can be interpreted as the stability of the dynamics.

In Tropical geometry scaling limits result in very different objects (dynamics) which are far from continuous processes.
Similar phenomena occur for instanton moduli spaces,
namely  compactification implies that at infinity of moduli spaces,
different spaces appear by bubbling-off instantons.
At any large scale the Chern classes are the same,  but in the limit
different Chern classes appear.

(2) 
The {\em Lotka-Volterra cell automaton} (LVCA) is given by the
equation:
$$u_{n+1}^{s+1} = u_n^{s+1} + \max\{L, u_{n+1}^s\} - \max\{ L, u_{n+2}^s\}.$$
It is shown in~\cite{K5} that there is a piecewise linear map $\varphi: {\mathbb R}^3 \to {\mathbb R}$
such that the corresponding dynamics 
$\Phi(\varphi): {\mathbb R}^{\infty} \to {\mathbb R}^{\infty}$ 
projects to the LVCA.
The corresponding complex dynamics is given by:
$$z_{n+1}^{s+1} = z_n^{s+1} \frac{t^L + z_{n+1}^s}{ t^L+z_{n+2}^s}~.$$
For $t= \infty$, the dynamics collapses and is described by the trivial equation
$z_{n+1}^{s+1} = z_n^{s+1}$.

\section{Intersection theory on the moduli spaces of curves and the matrix Airy function:}
We shall now give a mathematical application of our general considerations, namely to the Kontsevich-Witten model (KW)~\cite{K, Wit}, which originated in string theory.

KdV flows appear both in the physical and in the biological context. One of these arises from scale transformations of an automaton passing through Tropical geometry and the other shall be Witten's generating function, i.e. the tau-function, as it was shown by Kontsevich.
Thus  from our comparisons between physical and biological structures, it follows naturally  that the KW-model should have characteristics of an automaton.

So, let us give a commutative diagram for the Kontsevich model along the lines of our automation program. His approach to the Witten conjecture was based on three key objects: 
\begin{enumerate}
  \item One is to discretise the moduli space of Riemann surfaces 
via Strebel differentials and the thereto associated ribbon graphs.
  \item The second is to obtain the `main identity' by use of 
 Feynman diagram techniques.
  \item The third is to interpret Witten's generating function 
 via random matrices by use of the Airy function.
\end{enumerate}

Now let M${}^3$ or MMM stand for Miller-Morita-Mumford class and let us denote  discrete by `d' and ultra-discrete' by `ud' symbolically, such that   $\tau_d$ implies a discrete $\tau$-function etc.~.  Then our program of automation of the Kontsevich model 
yields the following commutative diagram:
$$ \begin{array}{ccccc} 
\text{\bf Automaton} &&  \text{\bf discrete~dynamics}  && \text{\bf continuous~dynamics}  \\
&&&& \\
	\tau_{ud}  & \leftarrow    &  \tau_d &   \rightarrow & \tau = \exp(F) \\
		\uparrow & & \uparrow && \uparrow \\
\text{Airy} \ \text{automaton} & \leftarrow  &  \text{discrete} \ \text{Airy} & \rightarrow &  \text{Airy \ function}  \\
	\uparrow & &&& \uparrow \\
(M^3~\text{class})_{ud} & \leftarrow &  (M^3~\text{class})_d & \rightarrow & \langle\tau_{d_1}, \dots, \tau_{d_n}\rangle  \\ 	
	\uparrow & & \uparrow && \uparrow \\
	{\mathfrak M}_{g,n}^{\text{ud}}  & \leftarrow & 
	{\mathfrak M}_{g,n}^{\text{d}} & \rightarrow &  {\mathfrak M}_{g,n}
\end{array}  $$
Let us remark that the discrete tau function $\tau_d$ can be obtained by the Miwa transform and that the middle objects on the vertical line have in part already been obtained by Chekhov~(\cite{chekhov}).

In~\cite{FK, FKT} we study another portion of the above diagram in detail, namely the one from the discrete MMM classes  $(M^3~\text{class})_d$
to Witten's generating function $F$. To do so, we shall combine three mathematical fields which have been developed independently, namely 
discrete integrable systems, moduli theory of Riemann surfaces, and the theory of mapping class groups.

The first two are connected by passing through 
discrete surface theory~\cite{BP}, whereas the last two 
through the group cohomology of the mapping class group.
So, let us denote the mapping class group of genus $g$ Riemann surfaces, which are finitely generated infinite groups, by $\Gamma$. 
Also, as it is known, the rational cohomology of the moduli space of Riemann surfaces is isomorphic to the group cohomology of $\Gamma$, i.e., 
$H^*({\frak M}_g : {\mathbb Q}) \cong H^*(\Gamma : {\mathbb{Q}})$.
Now, passing through these isomorphisms, the Poincar\'e dual of the 
Kontsevich cycles,  $a_n^*$ can be represented 
by elements of the group cohomology. Further, 
$H^*(\Gamma : {\mathbb Q})$ contains characteristic classes of 
surface bundles $\{\kappa_i\}_i$, which are called Miller-Morita-Mumford classes, with known explicit representation theoretic constructions of the cocycles for $M^3$ classes. 
Igusa~\cite{Igusa} obtained the leading terms of the Kontsevich cycles $a_n^*$ as:
$$
a_{(k_1^{n_1}  \dots k_r^{n_r})}^*
 = \Pi_{i=1}^r \frac{1}{n_i!} (2\frac{(2k_i+1)!}{(-1)^{k_i+1} k_i!})^{n_i}
( \tilde{\kappa}_{k_1})^{n_1}  \dots ( \tilde{\kappa}_{k_r})^{n_r}
  + \text{lower terms}.
$$

Now, our aim is to obtain a discrete dynamics, given by finite data,
for which the dynamical scaling limit is Witten's generating function $F$.
Another important point  for us is the construction of an automatic structure 
$(A,W,M_{g_i})$ on the mapping class group,
where $A=\{g_1, \dots, g_m\}$ is a set of  generators and $W$ is an automaton
with respect to $A$ (~\cite{Mo}).
Also, the discrete Miller-Morita-Mumford class
  $(A, W, M_{g_i}, \varphi)$ is equipped with an automatic structure and a map 
   $\varphi: A^{n+1}  \to \text{End}({\mathbb Z}^N)$, such that 
these sets  should  admit scaling limits which permit to 
 obtain parametrised group cocycles
$  e_n^t : \Gamma^{2n+1} \to {\mathbb Q}$ and its Taylor expansion 
$  \tilde{F} = \exp( \Sigma_n e_n^t  ) =
	  \sum_{(k)}\mu_{k_0, k_1, \dots} \prod^{\infty}_{t=0}\frac{t_i^{k_i}}{k_i!}~ $.
In the special case of genus $g=0$, the mapping class groups 
are isomorphic to the Braid groups divided by their centres.
In relation with integrable systems, 
Zabrodin~\cite{Zabrodin} found actions of the $R$-matrices, which will lead in our case to the construction of the $\varphi$ above.

{\bf Comments:}
(1)
On ${\mathfrak M}_{g,n}^{\text{comb.}} \to (\text{M}^3~\text{class})_d$.
For the construction of an automatic structure on the mapping class group, Mosher~\cite{Mo}  used triangulations on punctured Riemann surfaces, which will be related to another arrow in our previous diagram.
For surfaces of constant negative curvature,
the angle function of asymptotic directions satisfies the sine-Gordon equation.
On the other hand, according to the Thurston-Mumford approach to discretise the moduli spaces of Riemann surfaces, i.e., to give a cell decomposition, Kontsevich used Strebel differentials.
So, it would be interesting to consider the relation between the angle and the differentials, from the point of view of Schwarzian derivatives .

(2)
The moduli spaces of Riemann surfaces $\bar{\frak M}_{g,n}$ 
posses a discrete version $(\bar{\frak M}_{g,n})_{\text{disc}}$. Further, 
it is known that there is a symplectic structure $\omega$ on $\bar{\frak M}_{g,n}$.
Therefore, one can associate the symplectic Floer homology
$\text{HF}(\bar{\frak M}_{g,n} , \omega)$, which would 
give an idea how to construct a discrete version of symplectic Floer homology
directly from $(\bar{\frak M}_{g,n})_{\text{disc}}$.

(3) Yet another domain which naturally fits into our framework is quantum cosmology and the physics of black holes, cf.~\cite{tH}
\subsubsection*{Acknowledgements}
{The research of R.F. was partially supported by the Max-Planck-Gesellschaft.}

\end{document}